\documentclass[twocolumn, 10pt, journal,letterpaper]{IEEEtran}
\usepackage{amsmath}
\usepackage[pdftex]{graphicx}
\usepackage{graphicx}
\usepackage{pstricks}
\usepackage{amsfonts}
\usepackage{tabularx}
\newtheorem{Theorem}{Theorem}
\newtheorem{Lemma}{Lemma}
\newtheorem{Remark}{Remark}
\newtheorem{Proposition}{Proposition}

\newtheorem{Corollary}{Corollary}
\usepackage{amssymb}
\usepackage{multirow}
\usepackage{epsfig}
\usepackage{epstopdf}

\usepackage[noadjust]{cite} 
\usepackage[noend]{algpseudocode}
\usepackage{algorithm,algpseudocode,graphicx}

\begin{document}
\title{Decoding Orders for Securing Untrusted NOMA
 \\
}
\author{Sapna Thapar, Deepak Mishra, and Ravikant Saini
\thanks{S. Thapar and R. Saini are with the Department of Electrical Engineering, Indian Institute of Technology Jammu, Jammu $\&$ Kashmir 181 221, India (e-mail: thaparsapna25@gmail.com; ravikant.saini@iitjammu.ac.in). }
\thanks{D. Mishra is with the School of Electrical Engineering and Telecommunications, the University of New South Wales (UNSW) Sydney, NSW 2052, Australia (e-mail: d.mishra@unsw.edu.au).}
\thanks{This research work has been supported by the TCS RSP Fellowship.}}
 \maketitle
\begin{abstract}
This letter focuses on exploring a new decoding order to resolve the secrecy issue among untrusted non-orthogonal multiple access users. In this context, firstly, we identify the total number of possible decoding orders analytically. Then, we propose a decoding order strategy ensuring positive secrecy rate for all users, and obtain the number of feasible secure decoding orders based on this proposed strategy numerically. Thereafter, we present a suboptimal policy to avoid the computational complexity involved in finding the best secure decoding order. Numerical results illustrate that the suboptimal solution provides a performance gain of about $137\%$ over the relevant benchmark.

\textit{Index Terms$-$}
5G communications, physical layer security, non-orthogonal multiple access,  untrusted users.
\end{abstract}
\vspace{-4mm}
\section{Introduction}
Non-orthogonal multiple access (NOMA) is confronted with critical secrecy issues due to the broadcast nature of wireless transmission, and successive interference cancellation (SIC) based decoding at the receivers \cite{8114722}. Physical layer security (PLS) has recently emerged as a revolutionizing approach for security provisioning over wireless links in NOMA \cite{8509094}.

Utilizing the concept of PLS, existing works have mostly focused on securing NOMA against external eavesdroppers \cite{8509094}. Due to the inherent possibility of decoding other users' data while performing SIC, the design of secure NOMA when users are assumed to be untrusted is a challenge. Untrusted users' network is a practical and hostile situation where
users have no mutual trust, and each user focuses on securing its own data from all others, which leads to complex resource allocations \cite{7987792}, \cite{globecom}. In this context, assuming only far users as untrusted, sum secrecy rate of trusted near users is investigated in \cite{7833022} for a NOMA system. In \cite{basepaper} also, the far user is assumed as untrusted and secrecy outage probability of trusted near user is analyzed. However, near user has to decode far user's data to apply SIC, this results in a crucial security risk for the far user in case of untrusted near user. In this regard, secure decoding order for a two-user untrusted NOMA is proposed in \cite{globecom} which ensures positive secrecy rate for both the users. 

Compared to the two-user case \cite{globecom}, the selection of secure decoding order for multi-user untrusted scenario is a combinatorial and complicated problem. Besides, \cite{globecom} assumed an ideal setup with perfect SIC, which is not practical due to various implementation problems such as decoding error and complexity scaling \cite{8114722}, \cite{7881111}. \textit{To this end, considering imperfect SIC at receivers, we study secure decoding order selection problem for multi-user untrusted NOMA, which to the best of our knowledge, has not been investigated yet in the literature}. 

The key contributions are as follows: (1) A novel decoding order strategy for multi-user untrusted NOMA system is proposed. The key idea is to obtain positive secrecy rate at each user. (2) For a system having $N$ untrusted users, we analytically obtain the total number of possible decoding orders. Based on the proposed decoding order strategy, we present insight on the count of possible secure decoding orders that ensure positive secrecy rate to all users. (3) To reduce the computational complexity involved in exploring the best secure decoding order, we explore such favourable secure decoding orders that can provide higher secrecy rate for each user. We also offer a suboptimal decoding order policy. (4) Numerical results are provided to validate the proposed approach and quantify the achievable secrecy rate performance gain by the suboptimal solution over the relevant benchmark. 

%\textit{Notations}: We use bold upper (lower) letters to represent matrices (column vectors). $[\textbf{A}]_{u,v}$ stands for $(u, v)$-th element of matrix $\textbf{A}$, and $[\textbf{a}]_{u}$ indicates $u$-th element of vector $\textbf{a}$.

\section{Secure NOMA with Untrusted Users}\label{section2}
\subsection{System Model and NOMA Principle}\label{sys_model}
We consider the power-domain downlink NOMA system with a base station (BS) and $N$ untrusted users. The $n$-th user is denoted by $U_{n}$, where $n\in \mathcal{N} = \{1,2,...,N\}$. Each node in the system is equipped with a single antenna. The Rayleigh fading channel gain coefficient between BS and $U_{n}$ is denoted by $h_{n}$. The channel power gain $|h_{n}|^{2}$ follows exponential distribution with mean $\lambda_{n}=L_{c}d_{n}^{-e}$, where $L_{c}$, $e$, and $d_{n}$ denote path loss constant, path loss exponent, and distance from BS to $U_{n}$, respectively \cite{basepaper}. Without loss of generality, channel power gains are sorted such as $|h_{1}|^{2}>|h_{2}|^{2}>...>|h_{N}|^{2}$. %Based on channels' conditions between BS and users, $U_{1}$ and $U_{N}$ are considered as strongest and weakest users, respectively.
%The received signal at $U_{m}$, where $m \in \mathcal{N}$, can be given by \cite{basepaper}
%\begin{equation}
%y_{m} = h_{m}\sum_{n\in \mathcal{N}}\sqrt{P_{t}\alpha_{n}}x_{n} + w_{m},
%\end{equation}
%where $x_{n}$ is a unit power information signal meant for $U_{n}$, 
The BS superimposes users' information signals and broadcasts the generated mixture with total transmission power $P_{t}$. $\alpha_{n}$ denotes power allocation (PA) coefficient, i.e., the fraction of $P_{t}$ assigned to $U_{n}$ satisfying $\sum_{p\in \mathcal{N}}\alpha_{p}=1$. At receiver side, $U_{n}$ extracts the desired signal from the superimposed signal by cancelling the interfering signals through the SIC process. Without loss of generality, we assume  received additive white Gaussian noise for all users with zero mean and variance $\sigma^{2}$. We consider an imperfect SIC process where SIC could not be performed ideally due to practical limitations. Therefore, residual interference from imperfectly decoded signals exists while decoding later users. The residual interference can be expressed by a factor $\zeta$, $(0\leq\zeta\leq1)$, where $\zeta=0$ represents  perfect SIC while $\zeta=1$ refer to totally unsuccessful SIC \cite{7881111}.

\subsection{Proposed Decoding Order Representation}\label{DO}
While performing SIC, each user decodes its data as well as other users' data in a certain sequence. Thus, the collection of such sequences allocated for each user in the system can be defined by ``\textit{decoding order}'' of the system. According to conventional decoding order strategy \cite{8114722}, a stronger user first decodes data of all weaker users in ascending order of users' channel gain strength and then decodes its own data, and a weaker user never decodes the stronger user's data. Conversely, in the case of untrusted users, each user can decode data of itself as well as all other users. Note that SIC on the receivers’ side is a physical layer capability which enables the receiver to decode packets that arrive collectively. Therefore, each user can decode data of any user at any stage \cite{7833022}-\cite{globecom}, \cite{7343355} which results in a large number of decoding orders, which we will discuss in more detail in Section \ref{total_orders}. Considering these possibilities into account, let us represent the decoding order for an $N$-user system as an $N\times N$ matrix $\mathbf{D}_{o}$, where $o$ is an index indicating the $o$-th decoding order. The $m$-th column of matrix $\mathbf{D}_{o}$, where $m \in \mathcal{N}$, is represented by a column vector $\mathbf{d}_{m}$ of size $N\times1$, which demonstrates the SIC sequence followed by $U_{m}$. More specifically, $k$-th element of $\mathbf{d}_{m}$, i.e., $[\mathbf{d}_{m}]_{k}=n$ signifies that $U_{m}$ decodes data of $U_{n}$ at $k$-th stage, where $n, k \in \mathcal{N}$ and $[\mathbf{d}_{m}]_{1} \neq [\mathbf{d}_{m}]_{2} \hspace{0.1cm}\neq...\hspace{0.1cm}\neq [\mathbf{d}_{m}]_{N}$. 

\subsection{Achievable Secrecy Rates at Users in Untrusted NOMA}\label{secrecy_rates}
The signal-to-interference-plus-noise-ratio, $\gamma^{o}_{nm}$, when $U_{m}$ decodes data of $U_{n}$, where $m, n \in \mathcal{N}$, can be defined as 
\begin{equation}\label{sinr}
\gamma^{o}_{nm}= \frac{\alpha_{n} |h_{m}|^2}{\big(\zeta\sum_{i\in \mathcal{U}_{om}^{bn} }\alpha_{i}+\sum_{j\in \mathcal{U}_{om}^{an}}\alpha_{j}\big)|h_{m}|^2+\frac{1}{\rho_{t}}}. 
\end{equation}
$\mathcal{U}_{om}^{bn}$ denotes set of users' indices, $i$, which are decoded \textit{before} decoding of $U_{n}$ by $U_{m}$, where $i \in \mathcal{N} \setminus \{n\}$. Similarly, set $\mathcal{U}_{om}^{an}$ denotes those users' indices, $j$, which will be decoded \textit{after} decoding of $U_{n}$ by $U_{m}$, where $j\in \mathcal{N}\setminus \{n\}$. Thus, $\mathcal{U}_{om}^{bn}$ and $\mathcal{U}_{om}^{an}$ are disjoint sets such that $\mathcal{U}_{om}^{bn}\cap\mathcal{U}_{om}^{an}=\varnothing$ and $\mathcal{U}_{om}^{bn}\cup\mathcal{U}_{om}^{an}=\mathcal{N}\setminus\{n\}$. $\rho_{t}\stackrel{\Delta}{=}\frac{P_{t}}{\sigma^{2}}$ is BS transmit signal-to-noise ratio. The corresponding achievable data rate is given by 
\begin{equation} \label{info_rate}
R^{o}_{nm} = \log_{2}(1+\gamma^{o}_{nm}).
\end{equation}

Utilizing the concept of PLS, we investigate the achievable secrecy rate $R^{o}_{sn}$ at $U_{n}$, which is given by \cite{8509094}, \cite{7987792}
\begin{equation}\label{secrecy_rate}
R^{o}_{sn} = \{R^{o}_{nn} - \underset{m\in \mathcal{N} \setminus \{n\}}{\max} R^{o}_{nm}\}^+,
\end{equation}
where $\{\triangle\}^{+}=\max(0,\triangle)$. \textit{Note that the key idea of achieving positive secrecy rate at $U_{n}$ is to ensure that the condition $R^{o}_{nn}> R^{o}_{nm}$, simplified as $\gamma^{o}_{nn} > \gamma^{o}_{nm}$, must be satisfied.} 

\section{Decoding Order Strategy for Secure NOMA} 
%To design a secure NOMA system with untrusted users, here we focus on finding such a decoding order strategy which safeguards each user's confidentiality against each other.
\subsection{Possible Decoding Orders for Untrusted NOMA} \label{total_orders}
%We aim to design a secure NOMA system with untrusted users, in which each user's data can be secured from others. For this, 
Below we first identify the total number of possible decoding orders for an $N$-user untrusted NOMA system.
\begin{Proposition}\label{total_decoding_orders}
\textit{The total number of possible decoding orders for an $N$-user untrusted NOMA system is equal to $(N!)^{N}$.} 
\end{Proposition}
\begin{IEEEproof}
In an untrusted NOMA system, each user can decode its data as well as other users' data. Thus, for an $N$-user system, each user has a total of $N$ stages to decode data of all users. Considering the possibility to decode itself and other users' data at any stage, the number of permutations at each user is $N!$. As a result, we see that the count of total permutations for an $N$-user system, i.e., the number of possible decoding orders is $\underbrace{N! \times N! \hspace{0.1cm} ...\hspace{0.1cm} N!}_{N \text{times}} = (N!)^N$. 
\end{IEEEproof} 

Let us define a set of total possible decoding orders for an $N$-user system as $\mathcal{T}=\{\mathbf{D}_{o} | 1 \leq o \leq (N!)^{N} \}$. %Out of the total decoding orders, next, we show the incapability of that decoding order in providing a secure system which extends the conventional decoding order strategy in an untrusted scenario.

\subsection{Conventional NOMA Strategy in Untrusted Scenario}
Extending the conventional decoding order strategy \cite{8114722} in untrusted scenario, all users decode signals in ascending order of users' channel gain strength. Thus, $\gamma^{o}_{nm}$ can be written as
\begin{equation}
\gamma^{o}_{nm}= \frac{\alpha_{n} |h_{m}|^2}{\big(\zeta\sum_{i=n+1}^{N}\alpha_{i} + \sum_{j=1}^{n-1}\alpha_{j}\big)|h_{m}|^2+\frac{1}{\rho_{t}}}.
\end{equation}

Below we provide a key result on the inability of conventional NOMA strategy to secure an untrusted NOMA system.
\begin{Lemma}\label{conventional}
\textit{Utilizing the conventional decoding order strategy of NOMA in an untrusted scenario, the data of a weaker user cannot be secured from its respective stronger user.}
\end{Lemma}
\begin{IEEEproof}
To analyze the secrecy performance for a weaker user against the stronger user, we study secrecy rate $R^{o}_{sn}$ at $U_{n}$ against $U_{m}$ as defined in \eqref{secrecy_rate}, where $m<n$ and $m,n\in\mathcal{N}$. The required condition $\gamma^{o}_{nn}>\gamma^{o}_{nm}$ given in Section \ref{secrecy_rates} for positive $R^{o}_{sn}$ at $U_{n}$ gives $|h_{n}|^{2}>| h_{m}|^{2}$ which is infeasible as $|h_{m}|^{2}>| h_{n}|^{2}$ is considered. Thus, positive secrecy rate for a weaker user against the stronger user cannot be obtained.
\end{IEEEproof}

Next, we provide a result on the secrecy rate performance for stronger users against weaker users by Corollary \ref{A2}.
\begin{Corollary}\label{A2}
\textit{With conventional NOMA strategy, the stronger user's data is always safe from its respective weaker user.}
\end{Corollary}
\begin{IEEEproof}
Using \eqref{secrecy_rate}, we investigate $R^{o}_{sn}$ at $U_{n}$ against $U_{m}$, where $m>n$ and $m,n\in\mathcal{N}$. We find that $\gamma^{o}_{nn}>\gamma^{o}_{nm}$, required condition for positive $R^{o}_{sn}$, gives a feasible condition $|h_{n}|^{2}>| h_{m}|^{2}$, which ensures positive secrecy rate for $U_{n}$.
\end{IEEEproof}

\begin{Remark}
\textit{It can be easily inferred from the aforementioned results that by extending the conventional decoding order strategy in an untrusted scenario, only the strongest user's data is secured from all other users, and therefore, this strategy is not suitable for a secure NOMA system with untrusted users.} 
\end{Remark}

\subsection{Proposed Decoding Order Strategy for Secure NOMA}
 Considering the motive to secure each user's data from all others, below we propose a new decoding order strategy.
\begin{Theorem}\label{theorem}
\textit{If $U_{n}$ decodes its data at $k$-th stage and $U_{m}$ decodes data of $U_{n}$ at $k'$-th stage, where $m, n, k, k' \in \mathcal{N}$, $m<n$, and $k'< k$, then the weaker user's data can be safeguarded against the stronger user with a suitable constraint on PA .}
\end{Theorem}
\begin{IEEEproof}
We find $\gamma^{o}_{nn}$ and $\gamma^{o}_{nm}$ by using \eqref{sinr} to analyze the secrecy rate $R^{o}_{sn}$ defined in \eqref{secrecy_rate} for $U_{n}$ against $U_{m}$, where $m<n$ and $m,n\in\mathcal{N}$. For positive $R^{o}_{sn}$ at $U_{n}$, the required condition $\gamma^{o}_{nn}>\gamma^{o}_{nm}$ given in Section \ref{secrecy_rates} gives 
\begin{equation}\label{SINR}
\frac{\zeta\sum_{i\in \mathcal{U}_{om}^{bn} }\alpha_{i}+\sum_{j\in \mathcal{U}_{om}^{an}}\alpha_{j}+\frac{1}{\rho_{t}}}{\zeta\sum_{i\in \mathcal{U}_{on}^{bn} }\alpha_{i}+\sum_{j\in \mathcal{U}_{on}^{an}}\alpha_{j}+\frac{1}{\rho_{t}}}  > \frac{|h_{m}|^2}{|h_{n}|^2}.
\end{equation}
If $U_{n}$ decodes its data at $k$-th stage and $U_{m}$ decodes data of $U_{n}$ at $k'$-th stage, where $k'<k$ and $k, k' \in \mathcal{N}$, then it means $U_{m}$ decodes data of $U_{n}$ at least one stage before the stage of $U_{n}$ decoding itself. As a result, $\mathcal{U}_{om}^{bn}$ has at least one less interfering term in comparison to $\mathcal{U}_{on}^{bn}$, and $\mathcal{U}_{om}^{an}$ has at least one more interfering term compared to $\mathcal{U}_{on}^{an}$. Thus, due to more interfering terms in $\mathcal{U}_{om}^{an}$, we always find a PA condition on solving \eqref{SINR}. This result shows that we can obtain positive secrecy rate for the weaker user against the stronger user with a suitable PA constraint \cite{globecom}.
\end{IEEEproof} 

Next, a key result on the secrecy rate performance for stronger users against weaker users is given by Corollary \ref{A1}.
\begin{Corollary}\label{A1}
\textit{By using the proposed decoding order strategy stated in Theorem \ref{theorem}, the stronger user's data can also be secured from the respective weaker user.} 
\end{Corollary}
\begin{IEEEproof} 
For positive $R^{o}_{sn}$ at $U_{n}$ against $U_{m}$ when $m>n$, the required condition $\gamma^{o}_{nn}>\gamma^{o}_{nm}$ gives either a feasible condition $|h_{n}|^{2}>| h_{m}|^{2}$ or a suitable PA constraint. This result ensures that positive secrecy rate at $U_{n}$ can be obtained.
\end{IEEEproof}

Based on the proposed decoding order strategy given in Theorem 1, a set of decoding orders can be obtained which ensure positive secrecy rate for each user against all other users. This set is denoted by $\mathcal{S}$ and is termed as set of \textit{secure decoding orders}. Let us analytically define the set of these secure decoding orders for an $N$-user untrusted NOMA system as $\mathcal{S}=\{\mathbf{D}_{o} | 1 \leq o \leq (N!)^N, \text{and} \hspace{1mm} [\mathbf{d}_{n}]_{k}=n, [\mathbf{d}_{m}]_{k'}=n, m<n, k'<k, m, n, k, k' \in \mathcal{N}\}$. Thus, $\mathcal{S} \subset \mathcal{T}$. Note that since the motive is to secure data of each user from all others, we further focus on the secure decoding orders obtained based on the proposed decoding order strategy in Theorem 1.
\begin{Remark}\label{remarkk2}
\textit{It is obtained numerically that the count of secure decoding orders based on the proposed decoding order strategy for $2$, $3$, and $4$ users are $1$, $12$, and $3036$, respectively.}
\end{Remark}

%Note that only one secure decoding order exists for the case of two user NOMA, and the same has been proposed in \cite{globecom}.
\begin{Remark}
\textit{We have studied the system with $N\leq4$ due to two major reasons: first, the number of total decoding orders is enormous with more users due to which excessive computational complexity occurs in finding secure decoding orders and second, the number of users should not be too large in NOMA because it is an interference-limited system and implementation complexity increases at the transmitter side and receiver side with an increase in the number
of users \cite{ding2016impact}. However, the investigation can be extended for more users.}
\end{Remark}

\section{Low-Complexity Suboptimal Design}
After identifying set $\mathcal{S}$ of secure decoding orders, we focus on finding an optimal secure decoding order that can maximize the minimum secrecy rate among users. The corresponding
optimization problem over set $\mathcal{S}$ can be formulated as
\begin{align}\label{optimization_problem}
OP: \underset{ \mathbf{D}_{o} \in \mathcal{S}}{\max} \quad \underset{n \in \mathcal{N}}{\min} \quad R^{o}_{sn}, 
\quad &\text{s.t.} \quad C1: R^{o}_{sn} > 0.\nonumber
\end{align}

$OP$ is a complex combinatorial problem because a feasible PA is to be obtained for each decoding order that satisfies $C1$ and optimization is to be performed over the set $\mathcal{S}$ of secure decoding orders which keep increasing with $N$ as given in Remark \ref{remarkk2}. Therefore, to reduce this complexity, we select such favourable secure decoding orders from set $\mathcal{S}$ that can provide higher secrecy rate for each user by the approach given below.
\subsection{Favourable Secure Decoding Orders}
 \begin{Proposition}\label{efficient}
%\textit{The favourable secure decoding orders towards increasing the secrecy rate for each user are those in which each user decodes its data after decoding the data of all others.}
\textit{The favourable secure decoding orders providing relatively higher secure rate are those in which each user decodes its own data after decoding the data of all others.}
 \end{Proposition}
 \begin{IEEEproof}
For a decoding order $\mathbf{D}_{o}$, the secrecy rate $R^{o}_{sn}$ for each user $U_{n}$ defined in \eqref{secrecy_rate} can be increased by increasing $R^{o}_{nn}$, where $n \in \mathcal{N}$. From \eqref{info_rate}, we note that $R^{o}_{nn}$ increases with an increase in $\gamma^{o}_{nn}$, and a maximum value of $\gamma^{o}_{nn}$ is obtained when interference from other users is minimum. If $U_{n}$ decodes its data at the last stage, then $\mathcal{U}_{on}^{an} = \varnothing$ and $\mathcal{U}_{on}^{bn} = \mathcal{N}\setminus \{n\}$. As a result, a maximum value of $\gamma^{o}_{nn}$ is obtained since only the residual interference due to imperfect SIC remains in the denominator of $\gamma^{o}_{nn}$. Thus, decoding of own data at the last stage by each user can improve secrecy rate performance.  
\end{IEEEproof}

Let us represent a set of such favourable secure decoding orders in which each user decodes its data at the end as $\mathcal{L}$. For all remaining secure decoding orders of set $\mathcal{S}$ in which no user decodes its data at the last stage, we define a set $\mathcal{O}$. Here $\mathcal{L} \subset \mathcal{S}$, $\mathcal{O} \subset \mathcal{S}$, $\mathcal{L}\cap\mathcal{O} =\varnothing$ and $\mathcal{L}\cup\mathcal{O} = \mathcal{S}$. Since the achievable secrecy rate at each user is less in the secure decoding orders of set $\mathcal{O}$ in comparison to the secure decoding orders of set $\mathcal{L}$, we neglect the set $\mathcal{O}$ to reduce the computational complexity in finding the optimal solution. Next, an analytical insight on the count of favourable secure decoding orders is provided.

\begin{Proposition}\label{last_stage_decoding_orders}
\textit{For an $N$-user system, the total number of favourable secure decoding orders  is $((N-1)!)^{N}$.}
\end{Proposition}
\begin{IEEEproof}
In case of decoding own data at the last stage, each user has $(N-1)$ stages to decode other users' data. Thus, the number of permutations at each user $(N-1)!$. As a result, the count of favourable secure decoding orders for $N$-user system is $\underbrace{(N-1)! \times (N-1)! \hspace{0.1cm} ...\hspace{0.1cm} (N-1)!}_{N \text{ times}} = ((N-1)!)^N$.  
\end{IEEEproof}

From Proposition \ref{last_stage_decoding_orders}, we observe that secure decoding orders in set $\mathcal{L}$, are still in large number. Therefore, to avoid any computational complexity, we now design a suboptimal policy.  

\subsection{Suboptimal Decoding Order Policy}
Observing the dependence of secrecy rate performance on PA, we consider two PA schemes such as \textit{(i) lesser PA to the weaker user in comparison to the stronger user (LPWU),} and \textit{(ii) lesser PA to the stronger user in comparison to the weaker user (LPSU)}. %Taking these both PA schemes into account, 
Let us define PA coefficient $\alpha_{n}$ for $U_{n}$ as 
\begin{equation}
\alpha_{n}=\frac{1}{{(|h_{n}|^2)^\beta} \big(\sum_{p\in \mathcal{N}} \frac{1}{(|h_{p}|^2)^{\beta}}\big)},
\end{equation}
where $\beta$ is a real number $(-1\leq \beta \leq 1)$ with $\beta<0$ and $\beta>0$, respectively, representing $\textit{LPWU}$ and $\textit{LPSU}$ schemes. Considering both the PA schemes, our proposed suboptimal decoding order policy for an $N$-user system follows two steps:

$\bullet$ \textit{Step 1: For the first $(N-1)$ stages, each user decodes data of other users in the sequence of weakest to strongest user \text{(W-S)} for LPWU scheme, and in the order of strongest to weakest user \text{(S-W)} for LPSU scheme} $\rightarrow$ In the case of \textit{LPWU} scheme, the possibility of decoding weaker users' data by stronger users is higher due to their better channel gain and more PA. In such a way, the chance of decoding the weakest user's data by all other users is maximum. For this case, if the weakest user's data is decoded at the first stage by stronger users, then the achievable data rate decreases due to interference by other users. As a result, the secrecy rate for the weakest user improves. Considering this concept, we find that the (W-S) approach should be followed at all users. Similarly, \textit{LPSU} scheme represents the maximum PA to the weakest user. In this case, with a focus to increase the secrecy rate for stronger users, (S-W) approach should be followed.

$\bullet$ \textit{Step 2: At $N$-th stage each user decodes its data in both PA schemes} $\rightarrow$ As explained in Proposition \ref{efficient}, decoding own data at last stage by each user improves secrecy performance.

The detailed methodology is outlined in Algorithm \ref{Algo:AL1}. 
\begin{algorithm}[!htp]
{\small
\caption{\small Suboptimal decoding order policy.}\label{Algo:AL1}
\begin{algorithmic}[1]
\Require $N$, $\beta$, $\mathcal{N}$  
\Ensure suboptimal secure decoding order 
\State Define an $N \times N$ matrix $\mathbf{D}_{o}$ for $o$-th decoding order as explained in Section \ref{DO}
\For{each $m$-th column vector $\textbf{d}_{m}$ of matrix $\mathbf{D}_{o}$}
\State Obtain set $\mathcal{N}$ excluding the element $m$, such as $\mathcal{N} = \mathcal{N} \setminus \{\text{m}\}$
\For{each $q$-th element $t$ of set $\mathcal{N}$}
\State Set $[\textbf{d}_{m}]_{q} = t$
\EndFor
\State Set $[\textbf{d}_{m}]_{N} = m$ \hspace{.05cm} $\triangleright$ $m$-th user decodes its own data at end
\If{$\beta<0$} \hspace{2.7cm} $\triangleright$ \textit{LPWU} scheme - (W-S)
\State Sort the first $N-1$ elements of $\textbf{d}_{m}$ in descending order  
\ElsIf{$\beta>0$} \hspace{2.3cm} $\triangleright$ \textit{LPSU} scheme - (S-W)
\State Sort the first $N-1$ elements of $\textbf{d}_{m}$ in ascending order  
\EndIf
\For{each stage $k$ of matrix $\mathbf{D}_{o}$}
\State Set $[\mathbf{D}_{o}]_{k,m}=[\textbf{d}_{m}]_{k}$
\EndFor
\EndFor
\State return $\mathbf{D}_{o}$ \hspace{1.1cm} $\triangleright$ $\mathbf{D}_{o}$ as a suboptimal secure decoding order
\end{algorithmic}
}
\end{algorithm}

We denote the suboptimal decoding order for \textit{LPWU} and \textit{LPSU} schemes as $\widetilde {\mathbf{D}}$ and $\widehat{\mathbf{D}}$, respectively. Note that the  proposed algorithm is based on the sorting of elements. So, the computational complexity can be expressed as \textbf{O}(n log(n)).

\section{Numerical Results and Discussion}
In this work, users are distributed over a square field with length $l=500$m and BS is placed at its center. We have used $L_{c}=1$, $e=3$, and small scale fading is assumed to have Rayleigh distribution with mean value $1$. Besides, $\zeta=0.1$, $P_{t}=10$dBm and $\sigma^2=-90$dBm. Numerical results for each decoding order have been obtained in terms of the \textit{minimum secrecy rate} that can be ensured to each user. 

\begin{figure}[!t]
\centering
\includegraphics[scale=.33]{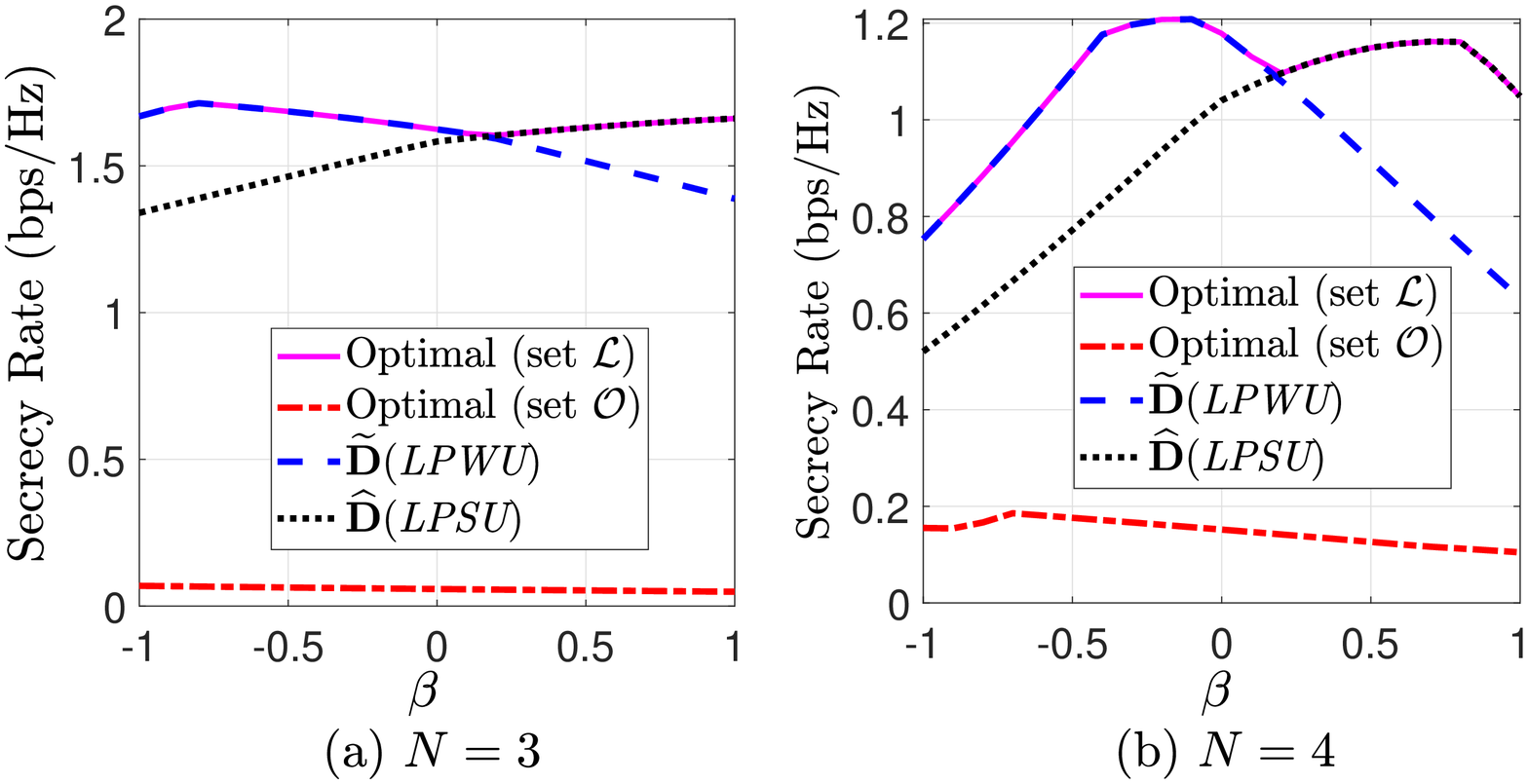}
\vspace{-8mm}
\caption{Validating the accuracy of proposed approach with a variation of secrecy rate performance for different PAs.}
\label{plot1}
\end{figure}

We first validate the performance of our proposed approach in Fig. \ref{plot1}. For this, the optimal secrecy rates achieved among secure decoding orders in set $\mathcal{L}$ and $\mathcal{O}$ are depicted for different PAs. The results reveal that the secure decoding orders in which all users decode their data at the last stage outperform that of other decoding orders. Fig. \ref{plot1} validates that $\widetilde {\mathbf{D}}$ and $\widehat{\mathbf{D}}$, respectively, are the suboptimal decoding orders for \textit{LPWU} and \textit{LPSU} schemes, except for some values of $\beta$. $\widetilde {\mathbf{D}}$ appears to work better than $\widehat{\mathbf{D}}$ until a certain point $\beta=0.2$ is reached, because, at the low value of $\beta$, the PA for weaker users is not enough to compensate the interference caused. Fig. \ref{plot1} also exhibits that the secrecy rate achieved through our proposed suboptimal solution match the optimal performance. Thus, the accuracy of suboptimal policy is about $90\%$.

\begin{figure}[!t]
\centering
\includegraphics[scale=.33]{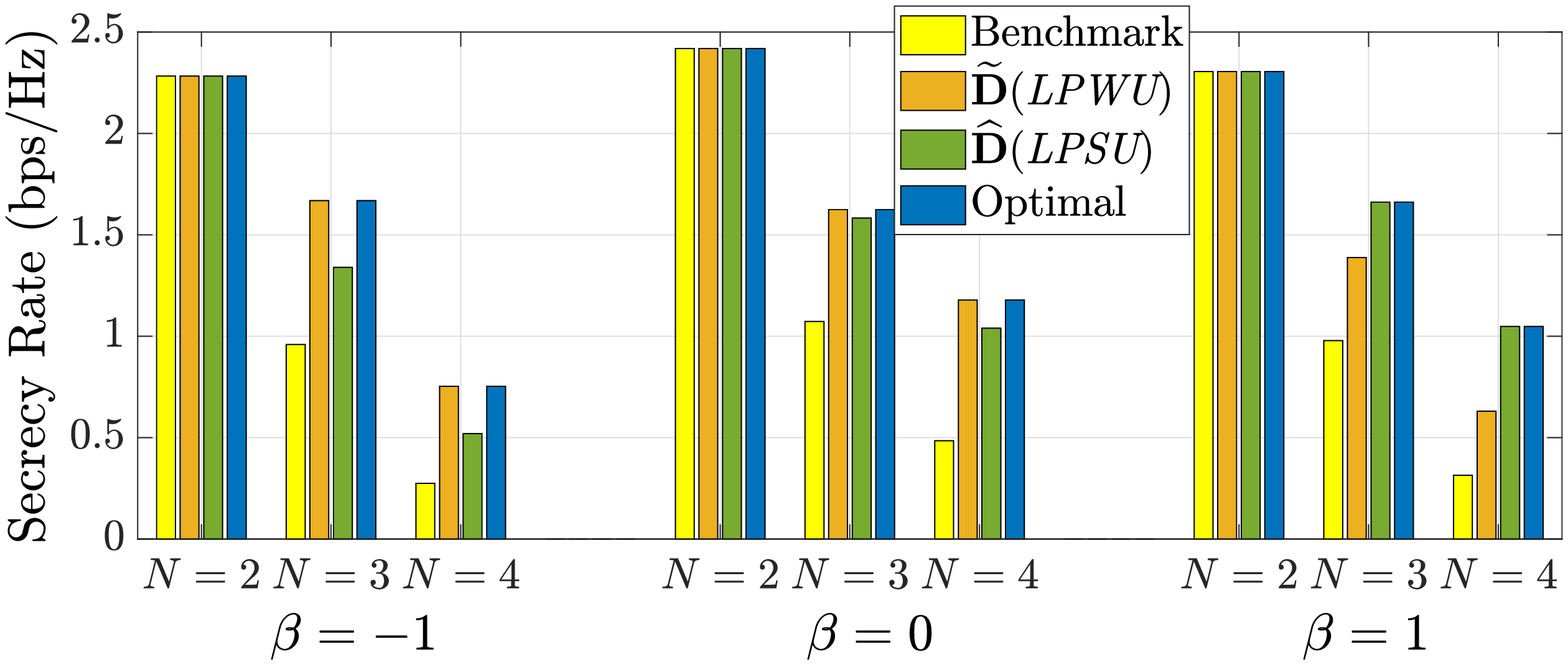}
\vspace{-8mm}
\caption{Comparison of the proposed suboptimal decoding order policy with the benchmark average secrecy rate of all secure decoding orders in set $\mathcal{S}$.}
\label{plot2}
\end{figure}

Next, Fig. \ref{plot2} presents the performance comparison of proposed suboptimal decoding orders with the benchmark. The benchmark scheme considered here is the average secrecy rate of all secure decoding orders in set $\mathcal{S}$. We observe that an average performance gain of about $137\%$ is achieved by suboptimal solution over the benchmark scheme. Fig. \ref{plot2} also shows that the secrecy rate performance degrades on increasing the number of users. This is because with an increase in the number of users, the achievable data rate at each user decreases due to interference by other users.

\section{Concluding Remarks}
We have proposed a decoding order strategy for solving secrecy issues in multi-user untrusted NOMA system. To avoid the computational complexity in finding the best decoding order out of the total possible secure decoding orders based on the proposed strategy, we have recommended a suboptimal policy. Significant improvement of $137\%$ in secrecy rate is achieved using the suboptimal solution over the benchmark. 

%\section*{Acknowledgement}
%This work has been supported by the TCS RSP Fellowship.

\bibliographystyle{IEEEtran}
\bibliography{ref}

\begin{thebibliography}{1}
\providecommand{\url}[1]{#1}
\csname url@rmstyle\endcsname
\providecommand{\newblock}{\relax}
\providecommand{\bibinfo}[2]{#2}
\providecommand\BIBentrySTDinterwordspacing{\spaceskip=0pt\relax}
\providecommand\BIBentryALTinterwordstretchfactor{4}
\providecommand\BIBentryALTinterwordspacing{\spaceskip=\fontdimen2\font plus
\BIBentryALTinterwordstretchfactor\fontdimen3\font minus
  \fontdimen4\font\relax}
\providecommand\BIBforeignlanguage[2]{{%
\expandafter\ifx\csname l@#1\endcsname\relax
\typeout{** WARNING: IEEEtran.bst: No hyphenation pattern has been}%
\typeout{** loaded for the language `#1'. Using the pattern for}%
\typeout{** the default language instead.}%
\else
\language=\csname l@#1\endcsname
\fi
#2}}

\bibitem{8114722}
Y.~{Liu}, Z.~{Qin}, M.~{Elkashlan}, Z.~{Ding}, A.~{Nallanathan}, and
  L.~{Hanzo}, ``Nonorthogonal multiple access for {5G} and beyond,''
  \emph{Proc. IEEE}, vol. 105, no.~12, pp. 2347--2381, Dec. 2017.

\bibitem{8509094}
J.~M. {Hamamreh}, H.~M. {Furqan}, and H.~{Arslan}, ``Classifications and
  applications of physical layer security techniques for confidentiality: A
  comprehensive survey,'' \emph{IEEE Commun. Surveys Tuts.}, vol.~21, no.~2,
  pp. 1773--1828, Secondquarter 2019.

\bibitem{7987792}
R.~Saini, D.~Mishra, and S.~De, ``Utility regions for {DF} relay in
  {OFDMA}-based secure communication with untrusted users,'' \emph{IEEE Commun.
  Lett.}, vol.~21, no.~11, pp. 2512--2515, Nov. 2017.

\bibitem{globecom}
S.~{Thapar}, D.~{Mishra}, and R.~{Saini}, ``Novel outage-aware {NOMA} protocol
  for secrecy fairness maximization among untrusted users,'' \emph{IEEE Trans.
  Veh. Technol.}, vol.~69, no.~11, pp. 13\,259--13\,272, 2020.

\bibitem{7833022}
Y.~{Li}, M.~{Jiang}, Q.~{Zhang}, Q.~{Li}, and J.~{Qin}, ``Secure beamforming in
  downlink {MISO} nonorthogonal multiple access systems,'' \emph{IEEE Trans.
  Veh. Technol.}, vol.~66, no.~8, pp. 7563--7567, Aug. 2017.

\bibitem{basepaper}
B.~M. {ElHalawany} and K.~{Wu}, ``Physical-layer security of {NOMA} systems
  under untrusted users,'' in \emph{Proc. IEEE GLOBECOM}, United Arab Emirates,
  Dec. 2018, pp. 1--6.

\bibitem{7881111}
H.~{Sun}, B.~{Xie}, R.~Q. {Hu}, and G.~{Wu}, ``Non-orthogonal multiple access
  with {SIC} error propagation in downlink wireless {MIMO} networks,'' in
  \emph{Proc. IEEE VTC-Fall}, Montreal, Canada, Sep. 2016, pp. 1--5.

\bibitem{7343355}
X.~{Chen}, A.~{Beiijebbour}, A.~{Li}, H.~{Jiang}, and H.~{Kayama},
  ``Consideration on successive interference canceller {(SIC)} receiver at
  cell-edge users for non-orthogonal multiple access {(NOMA)} with {SU-MIMO},''
  in \emph{Proc. PIMRC}, Hong Kong, China, Aug. 2015, pp. 522--526.

\bibitem{ding2016impact}
Z.~Ding, P.~Fan, and H.~V. Poor, ``Impact of user pairing on 5{G} nonorthogonal
  multiple-access downlink transmissions,'' \emph{IEEE Trans. Veh. Technol.},
  vol.~65, no.~8, pp. 6010--6023, Aug. 2016.

\end{thebibliography}

\end{document}